\documentclass[twocolumn,twoside,slac_two]{revtex4}
\usepackage{graphicx}
\usepackage{fancyhdr}
\begin{document}

\title{Hot topics from Belle experiment}

%

\author{K. Ikado}
\affiliation{Nagoya University, Nagoya, 464-8602, Japan}

\begin{abstract}
We present the first evidence of the decay $B^{-}\rightarrow\tau^{-}\bar{\nu}_{\tau}$, 
using $414~\textrm{fb}^{-1}$ of data collected at the $\Upsilon(4S)$ resonance 
with the Belle detector at the KEKB asymmetric-energy $e^{+}e^{-}$ collider. 
Events are tagged by fully reconstructing one of the $B$ mesons in hadronic modes.
We detect the signal with a significance of $4.0$ standard deviations including systematics, 
and measure the branching fraction to be 
${\cal B}(B^{-}\rightarrow\tau^{-}\bar{\nu}_{\tau}) = 
 (1.06^{+0.34}_{-0.28}(\mbox{stat})^{+0.22}_{-0.25}(\mbox{syst}))\times 10^{-4}$.
We also report results based on $1.86\mbox{fb}^{-1}$ data 
collected by the Belle detector at the $\Upsilon(5S)$ resonance. 
Several exclusive $B_{s}$ decays $B_{s}\rightarrow D_{s}^{(*)+}\pi^{-}(\rho^{-})$ and 
$B_{s}\rightarrow J/\psi\phi(\eta)$ are studied.
The $B_{s}$ meson production is found to proceed predominantly through the creation of 
$B_{s}^{*}\bar{B}_{s}^{*}$ pairs. 
Upper limits on 
$B_{s}\rightarrow K^+K^-$, $B_{s}\rightarrow\phi\gamma$, $B_{s}\rightarrow\gamma\gamma$
and $B_{s}\rightarrow D_{s}^{(*)+}D_{s}^{(*)-}$ decays are also reported.

\end{abstract}

\maketitle

\thispagestyle{fancy}


\section{Introduction}
In the Standard Model (SM), the purely leptonic decay 
$B^{-}\rightarrow\tau^{-}\bar{\nu}_{\tau}$
proceeds via annihilation of $b$ and $\overline{u}$ quarks to a $W^-$ boson.
It provides a direct determination of the product of the $B$ meson decay 
constant $f_B$ and the magnitude of the
Cabibbo-Kobayashi-Maskawa (CKM) matrix element $|V_{ub}|$.
The branching fraction is given by
\begin{equation}
 \label{eq:BR_B_taunu}
{\cal B}(B^{-}\rightarrow\tau^{-}\bar{\nu}_{\tau}) 
= \frac{G_{F}^{2}m_{B}m_{\tau}^{2}}{8\pi}\left(1-\frac{m_{\tau}^{2}}
{m_{B}^{2}}\right)^{2}f_{B}^{2}|V_{ub}|^{2}\tau_{B},
\end{equation}
where $G_F$ is the Fermi coupling constant, 
$m_{B}$ and $m_{\tau}$ are the $B$ and $\tau$ masses, respectively, 
and $\tau_B$ is the $B^-$ lifetime~\cite{Eidelman:2004wy}.
Physics beyond the SM, such as supersymmetry or two-Higgs doublet models,
could modify ${\cal B}(B^{-}\rightarrow\tau^{-}\bar{\nu}_{\tau})$ through
the introduction of a charged Higgs boson~\cite{Hou:1992sy}.
Purely leptonic $B$ decays have not been observed in past experiments.
The most stringent upper limit on $B^{-}\rightarrow\tau^{-}\bar{\nu}_{\tau}$
comes from the BaBar experiment: 
${\cal B}(B^{-}\rightarrow\tau^{-}\bar{\nu}_{\tau}) < 2.6 \times 10^{-4}$
(90\% C.L.)~\cite{Aubert:2005}.

The possibility to study decays of $B_{s}$ at very high luminosity $e^+e^-$ colliders running
at the energy of the $\Upsilon(5S)$ resonance has been discussed in several theoretical
papers~\cite{Falk:2000ga,Atwood:2001js}. 
The first data at the $\Upsilon(5S)$ were taken many years ago at CESR~\cite{Lovelock:1985nb,Besson:1984bd,Lee-Franzini:1990gy},
but the collected data sample was not enough to observe a $B_{s}$ signal. 
In 2003, the CLEO experiment collected $0.42~\mbox{fb}^{-1}$ at the $\Upsilon(5S)$
and observed some evidence for $B_{s}$ meson production in both inclusive and exclusive modes. 
However, simple calculations assuming an approximate $SU(3)$ symmetry indicate that many
interesting $B_{s}$ measurements require a data sample of at least $20~\mbox{pb}^{-1}$, 
which can be collected by $B$ Factories in the future. 
To test the experimental feasibility of such measurements, a data sample of $1.86~\mbox{fb}^{-1}$
was recently taken with the Belle detector at the center-of-mass (CM) energy corresponding to the mass
of the $\Upsilon(5S)$ resonance.
This data sample is more than four times larger than the CLEO dataset
at the $\Upsilon(5S)$.

The Belle detector is a large-solid-angle
magnetic spectrometer consisting of a silicon vertex detector,
a $50$-layer central drift chamber (CDC), a system of aerogel threshold
$\check{\textrm{C}}$erenkov counters (ACC), time-of-flight scintillation
counters (TOF), and an electromagnetic calorimeter comprised of
CsI(Tl) crystals (ECL)
located inside a superconducting solenoid coil that provides a $1.5$ T
magnetic field. An iron flux-return located outside of the coil is
instrumented to identify $K_{L}^{0}$ and muons.
The detector is described in detail elsewhere~\cite{belle_detector:2003}.

\section{Evidence of the Purely Leptonic Decay \boldmath$B^{-}\rightarrow\tau^{-}\bar{\nu}_{\tau}$}

We use a $414~\textrm{fb}^{-1}$ data sample containing 
$447\times 10^{6}$ $B$ meson pairs collected with the Belle detector
at the KEKB asymmetric-energy $e^{+}e^{-}$ ($3.5$ on $8$ GeV) collider
operating at the $\Upsilon(4S)$ resonance ($\sqrt{s} = 10.58$ GeV). 

We use a detailed Monte Carlo (MC) simulation, which fully describes the
detector geometry and response based on GEANT~\cite{GEANT}, to determine
the signal selection efficiency and to study the background.
In order to reproduce effects of beam background, data taken with random
triggers for each run period are overlaid on simulated events. 
The $B^{-}\rightarrow\tau^{-}\bar{\nu}_{\tau}$ signal decay is generated 
by the EvtGen package~\cite{EvtGen}.
To model the background from $e^+e^- \to B\overline{B}$ and continuum 
$q\overline{q}~(q = u, d, s, c)$ production processes, large 
$B\overline{B}$ and $q\overline{q}$ MC samples 
corresponding to about twice the data sample
are used.
We also use MC samples for rare $B$ decay processes, such as charmless 
hadronic, radiative, electroweak decays and $b \to u$ semileptonic decays.

We fully reconstruct one of the $B$ mesons in 
the event, referred to hereafter as the tag side ($B_{\rm tag}$), 
and compare properties of the remaining particle(s), referred to as the 
signal side ($B_{\rm sig}$), to those expected for signal and background.
The method allows us to suppress strongly the combinatorial background 
from both $B\overline{B}$ and continuum events.
In order to avoid experimental bias, 
the signal region in data is not looked at until the event 
selection criteria are finalized.

The $B_{\rm tag}$ candidates are reconstructed in the following decay modes: 
$B^{+} \rightarrow \overline{D}{}^{(*)0} \pi^{+}$, 
$\overline{D}{}^{(*)0}\rho^{+}$, 
$\overline{D}{}^{(*)0}a_{1}^{+}$ 
and $\overline{D}{}^{(*)0}D_{s}^{(*)+}$.
The $\overline{D}$ mesons are reconstructed as 
$\overline{D}{}^{0}\rightarrow K^{+}\pi^{-}$, $K^{+}\pi^{-}\pi^{0}$,
$K^{+}\pi^{-}\pi^{+}\pi^{-}$, $K_{S}^{0}\pi^{0}$, $K_{S}^{0}\pi^{-}\pi^{+}$,
$K_{S}^{0}\pi^{-}\pi^{+}\pi^{0}$ and $K^{-}K^{+}$, and
the $D_{s}^{+}$ mesons are reconstructed as 
$D_{s}^{+}\rightarrow K_{S}^{0}K^{+}$ and $K^{+}K^{-}\pi^{+}$.
The $\overline{D}{}^{*0}$ and $D_{s}^{*+}$ mesons are reconstructed in
$\overline{D}{}^{*0} \to \overline{D}{}^0 \pi^0, \overline{D}{}^0 \gamma$,
and $D_{s}^{*+} \to D_{s}^{+} \gamma$ modes.  
The selection of $B_{\rm tag}$ candidates is based on the 
beam-constrained mass $M_{\rm bc}\equiv\sqrt{E_{\rm beam}^{2} - p_{B}^{2}}$
and the energy difference $\Delta E\equiv E_{B} - E_{\rm beam}$.
Here, $E_{B}$ and $p_{B}$ are the reconstructed energy and momentum
of the $B_{\rm tag}$ candidate in the $e^+e^-$ center-of-mass system,
and $E_{\rm beam}$ is the beam energy in the CM frame.
The selection criteria for $B_{\rm tag}$ are defined as
$M_{\rm bc}>5.27~\mbox{GeV}/c^{2}$ and $-80~\mbox{MeV} <\Delta E< 60~\mbox{MeV}$.
If an event has multiple $B_{\rm tag}$ candidates, we choose the one having
the smallest $\chi^{2}$ based on deviations from the nominal values of 
$\Delta E$, the $D$ candidate mass, and the $D^{*} - D$ mass difference if
applicable.

In the events where a $B_{\rm tag}$ is reconstructed, we search for decays
of $B_{\rm sig}$ into a $\tau$ and a neutrino. 
Candidate events are required to have one or three charged track(s) on the
signal side with the total charge being opposite to that of $B_{\rm tag}$.
The $\tau$ lepton is identified in the five decay modes,
$\mu^{-}\bar{\nu}_{\mu}\nu_{\tau}$,
$e^{-}\bar{\nu}_{e}\nu_{\tau}$, 
$\pi^{-}\nu_{\tau}$,
$\pi^{-}\pi^{0}\nu_{\tau}$ and 
$\pi^{-}\pi^{+}\pi^{-}\nu_{\tau}$,
which taken together correspond to $81\%$ of all $\tau$ decays~\cite{Eidelman:2004wy}.
The muon, electron and charged pion candidates are selected based on
information from particle identification devices.
The leptons are selected with requirements that have efficiencies greater than 90\% 
for both muons and electrons in the momentum region above 1.2 GeV/$c$, and 
misidentification rates of less than 0.2\%(1.5\%) for electrons (muons) in 
the same momentum region. 
Kaon candidates are rejected for all charged tracks on the signal side.
The $\pi^0$ candidates are reconstructed by requiring the invariant mass of two
$\gamma$'s to satisfy $|M_{\gamma\gamma}-m_{\pi^0}| < 20~\mbox{MeV}/c^{2}$.
For all modes except $\tau^{-}\rightarrow\pi^{-}\pi^{0}\nu_{\tau}$, we reject events with 
$\pi^{0}$ mesons on the signal side.
All the selection criteria have been optimized to achieve the highest sensitivity in MC.

The most powerful variable for separating signal and background is the 
remaining energy in the ECL, denoted as $E_{\rm ECL}$, which is sum of
the energy of photons that are not associated with either the 
$B_{\rm tag}$ or the $\pi^{0}$ candidate from the 
$\tau^{-}\rightarrow \pi^{-}\pi^{0}\nu_{\tau}$ decay.
For signal events, $E_{\rm ECL}$ must be either zero or small value 
arising from beam background hits, therefore, signal events peak at 
low $E_{\rm ECL}$.
On the other hand, background events are distributed toward higher 
$E_{\rm ECL}$ due to the contribution from additional neutral clusters.

The $E_{\rm ECL}$ signal region is optimized for each $\tau$ decay mode based
on the MC simulation, and is defined by $E_{\rm ECL} < 0.2~\mbox{GeV}$ for the
$\mu^{-}\bar{\nu}_{\mu}\nu_{\tau}$, $e^{-}\bar{\nu}_{e}\nu_{\tau}$ and
$\pi^{-}\nu_{\tau}$ modes, and $E_{\rm ECL} < 0.3~\mbox{GeV}$ for
the $\pi^{-}\pi^{0}\nu_{\tau}$ and $\pi^{-}\pi^{+}\pi^{-}\nu_{\tau}$
modes. 
The $E_{\rm ECL}$ sideband region is defined by $0.4~\mbox{GeV} < E_{\rm ECL} < 1.2$ GeV 
for the $\mu^{-}\bar{\nu}_{\mu}\nu_{\tau}$, $e^{-}\bar{\nu}_{e}\nu_{\tau}$ and
$\pi^{-}\nu_{\tau}$ modes, and by $0.45~\mbox{GeV} < E_{\rm ECL} < 1.2$ GeV for
the $\pi^{-}\pi^{0}\nu_{\tau}$ and $\pi^{-}\pi^{+}\pi^{-}\nu_{\tau}$ modes.
Table~\ref{tab:signal_yields} shows the number of events found in the sideband 
region for data ($N_{\rm side}^{\rm obs}$) and for the background MC simulation 
($N_{\rm side}^{\rm MC}$) scaled to the equivalent integrated luminosity in data.
Their good agreement for each $\tau$ decay mode indicates the validity of the
background MC simulation.
Table~\ref{tab:signal_yields} also shows the number of the background MC events 
in the signal region ($N_{\rm sig}^{\rm MC}$).  

\begin{figure}
\begin{center}
 \includegraphics[width=80mm]{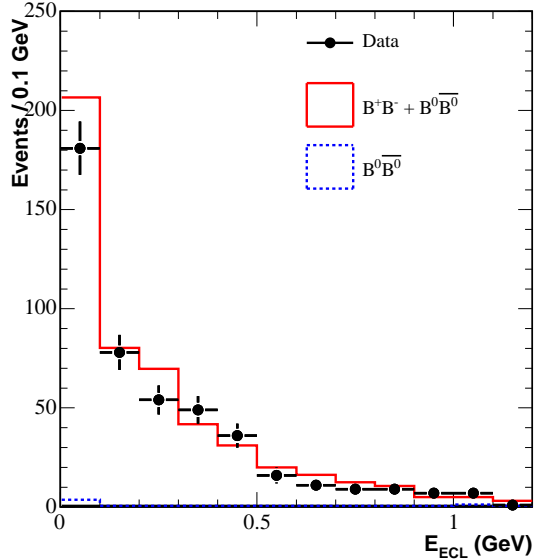}
\caption{$E_{\rm ECL}$ distribution for the control sample of doubly tagged events, where one $B$ is fully
  reconstructed and the other $B$ is reconstructed as $B^{-} \rightarrow D^{*0}\ell^{-}\bar{\nu}$. 
  The dots with errors indicate the data. The solid histogram represents the distribution as
  deduced background from 
  $B\overline{B}$ MC ($B^+B^- + B^0\overline{B}{}^0$), and the dashed histogram shows the contribution 
  from $B^0\overline{B}{}^0$ events.}
    \label{fig:controlsample}
\end{center}   
\end{figure}

In order to validate the $E_{\rm ECL}$ simulation, we use a control sample
of events (double tagged events), where the $B_{\rm tag}$ is fully reconstructed 
as described above and $B_{\rm sig}$ is reconstructed in the decay chain, 
$B^{-} \rightarrow D^{*0}\ell^{-}\bar{\nu}$ ($D^{*0}\rightarrow D^{0}\pi^{0}$),
followed by $D^0 \to K^- \pi^+$ or $K^- \pi^- \pi^+ \pi^+$
where $\ell$ is a muon or electron.
The sources affecting the $E_{\rm ECL}$ distribution in the control sample 
are similar to those affecting the $E_{\rm ECL}$ distribution in the signal 
MC simulation.
Figure~\ref{fig:controlsample} shows the $E_{\rm ECL}$ distribution in the
control sample for data and the MC simulation scaled to 
equivalent integrated luminosity in data.
Their agreement demonstrates the validity of the $E_{\rm ECL}$ simulation 
in the signal MC.

\begin{figure}
\begin{center}
 \includegraphics[width=80mm]{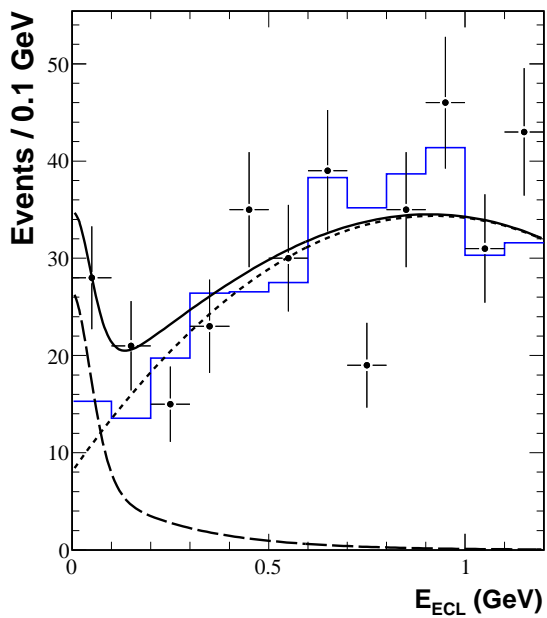} 
\caption{$E_{\rm ECL}$ distributions in the data after
	all selection requirements except the one on $E_{\rm ECL}$ 
	have been applied. The data and background MC samples are represented by the points with error bars
	and the solid histogram, respectively. 
	The solid curve shows the result of the fit with the sum of the signal shape (dashed) and background shape
	(dotted).
}
    \label{ecl_opened}
\end{center}   
\end{figure}

After finalizing the signal selection criteria, the signal region is examined.
Figure~\ref{ecl_opened} shows the obtained $E_{\rm ECL}$ distribution
when all $\tau$ decay modes are combined.
One can see a significant excess of events in the $E_{\rm ECL}$ signal region
below $E_{\rm ECL}< 0.25$ GeV.
Table~\ref{tab:signal_yields} shows the number of events observed in the 
signal region ($N_{\rm obs}$) for each $\tau$ decay mode.
For the events in the signal region, we verify that the distributions of the 
event selection variables other than $E_{\rm ECL}$, such as $M_{\rm bc}$ and
$p_{\rm miss}$, are consistent with the sum of the signal and background
distributions expected from MC.

We deduce the final results by fitting the obtained $E_{\rm ECL}$ 
distributions to the sum of the expected signal and background shapes.
Probability density functions (PDFs) for the signal $f_s(E_{\rm ECL})$ and
for the background $f_b(E_{\rm ECL})$ are constructed for each $\tau$ decay
mode from the MC simulation.
The signal PDF is modeled as the sum of a Gaussian function, centered at
$E_{\rm ECL} = 0$, and an exponential function.
The background PDF, as determined from the MC simulation, is parametrized 
by a second-order polynomial.
The PDFs are combined into an extended likelihood function,
\begin{equation}
{\cal L} = \frac{e^{-(n_{s}+n_{b})}}{N!}\prod_{i=1}^{N}(n_{s}f_{s}(E_{i})+n_{b}f_{b}(E_{i})),
\end{equation}
where $E_{i}$ is the $E_{\rm ECL}$ in the $i$th event, $N$ is the total number 
of events in the data, and $n_{s}$ and $n_{b}$ are the signal yield and background 
yield to be determined by the fit.
To combine likelihood functions of the five decay modes, we multiply 
the likelihood functions to produce the combined likelihood 
(${\cal L}_{\rm com} = \prod_{j=1}^{5} {\cal L}_{j}$).
The results are listed in Table~\ref{tab:signal_yields}.
The number of signal events in the signal region deduced from the fit ($N_{\rm s}$) is $21.2^{+6.7}_{-5.7}$
when all $\tau$ decay modes are combined.
Table~\ref{tab:signal_yields} also gives the number of background events 
in the signal region deduced from the fit ($N_{\rm b}$), which is 
consistent with the expectation from the background MC simulation 
($N_{\rm sig}^{\rm MC}$). 

The branching fractions are calculated as
${\cal B} = N_{s}/(2\cdot\varepsilon\cdot N_{B^{+}B^{-}})$
where $N_{B^{+}B^{-}}$ is the number of $\Upsilon(4S)\rightarrow B^{+}B^{-}$ 
events, assumed to be half of the number of produced $B$  meson pairs. 
The efficiency is defined as 
$\varepsilon = \varepsilon^{\rm tag}\times\varepsilon^{\rm sel}$,
where $\varepsilon^{\rm tag}$ is the tag reconstruction efficiency for events with 
$B^{-}\rightarrow\tau^{-}\bar{\nu}_{\tau}$ decays on the signal side, determined by MC
to be $0.136\pm 0.001({\rm stat})\%$,  and $\varepsilon^{\rm sel}$ is the event selection 
efficiency listed in Table~\ref{tab:signal_yields}, as 
determined by the ratio of the number of events surviving all of the 
selection criteria including the $\tau$ decay branching fractions over the number of fully 
reconstructed $B^{\pm}$.
When all $\tau$ decay modes are combined
we obtain a branching fraction of $(1.06^{+0.34}_{-0.28}) \times 10^{-4}$.
The branching fraction for each $\tau$ decay mode is consistent within  
error as shown in Table~\ref{tab:signal_yields}.

\begin{table*}
  \renewcommand{\baselinestretch}{1.3}
    \begin{tabular}{|c|c|c|c|c|c|c|c|c|c|} \hline
&$N_{\rm side}^{\rm obs}$ &$N_{\rm side}^{\rm MC}$ &$N_{\rm sig}^{\rm MC}$ &$N_{\rm obs}$ &$N_{\rm s}$ &$N_{\rm b}$
&$\varepsilon^{\rm sel}(\%)$  &${\cal B}(10^{-4})$  &$\Sigma$\\\hline
$\mu^{-}\bar{\nu}_{\mu}\nu_{\tau}$ &$96$   &$94.2\pm 8.0$   &$9.4\pm 2.6$  &$13$ &$5.4^{+3.2}_{-2.2}$  &$9.1^{+0.2}_{-0.1}$ &$8.88\pm 0.05$   &$1.01^{+0.59}_{-0.41}$ &$2.0\sigma$\\\hline
$e^{-}\bar{\nu}_{e}\nu_{\tau}$     &$93$   &$89.6\pm 8.0$   &$8.6\pm 2.3$  &$12$ &$3.9^{+3.5}_{-2.5}$  &$9.2^{+0.2}_{-0.2}$ &$8.18\pm 0.05$   &$0.79^{+0.71}_{-0.49}$ &$1.3\sigma$\\\hline
$\pi^{-}\nu_{\tau}$                &$43$   &$41.3\pm 6.2$   &$4.7\pm 1.7$  &$9$  &$3.4^{+2.6}_{-1.6}$  &$4.0^{+0.2}_{-0.1}$ &$5.79\pm 0.04$   &$0.96^{+0.74}_{-0.46}$ &$1.9\sigma$\\\hline
$\pi^{-}\pi^{0}\nu_{\tau}$         &$21$   &$23.3\pm 4.7$   &$5.9\pm 1.9$  &$11$ &$6.2^{+3.9}_{-2.7}$  &$4.2^{+0.3}_{-0.3}$ &$8.32\pm 0.08$   &$1.23^{+0.77}_{-0.53}$ &$2.3\sigma$\\\hline
$\pi^{-}\pi^{+}\pi^{-}\nu_{\tau}$  &$21$   &$18.5\pm 4.1$   &$4.2\pm 1.6$  &$9$  &$3.1^{+3.1}_{-2.6}$  &$3.7^{+0.3}_{-0.2}$ &$1.75\pm 0.03$   &$2.99^{+3.01}_{-2.49}$ &$1.2\sigma$\\
\hline
Combined                           &$274$  &$266.9\pm 14.3$ &$32.8\pm 4.6$ &$54$ &$21.2^{+6.7}_{-5.7}$ &$30.2^{+0.5}_{-0.4}$ &$32.92\pm 0.12$  &$1.06^{+0.34}_{-0.28}$   &$4.0\sigma$\\

\hline
    \end{tabular}
    \caption{The number of observed events in data in the sideband region $(N_{\rm side}^{\rm obs})$,
      number of background MC events in the sideband region $(N_{\rm side}^{\rm MC})$ and the 
      signal region $(N_{\rm sig}^{\rm MC})$,
      number of observed events in data in the signal region $(N_{\rm obs})$, 
      number of signal $(N_{\rm s})$ and background $(N_{\rm b})$ in the signal region determined by the fit, 
      signal selection efficiencies $(\varepsilon^{\rm sel})$, 
      extracted branching fraction $({\cal B})$ for $B^{-}\rightarrow\tau^{-}\bar{\nu}_{\tau}$.
      The listed errors are statistical 
      only. The last column gives the significance of the signal including 
      the systematic uncertainty in the signal yield ($\Sigma$).}
   \label{tab:signal_yields}
\end{table*}

Systematic errors for the measured branching fraction are associated with 
the uncertainties in the  number of $B^{+}B^{-}$, signal yields and  
efficiencies.
The total fractional uncertainty of the combined measurement is 
$^{+20.5}_{-24.0}\%$, 
and we measure the branching fraction to be
$$
{\cal B}(B^{-}\rightarrow\tau^{-}\bar{\nu}_{\tau}) 
= (1.06^{+0.34}_{-0.28}(\mbox{stat})^{+0.22}_{-0.25}(\mbox{syst}))\times 10^{-4}.
$$
The significance is $4.0\sigma$ when all $\tau$ decay modes are combined, 
where the significance is defined as 
$\Sigma = \sqrt{-2\ln({\cal L}_{0}/{\cal L}_{\rm max})}$,
where ${\cal L}_{\rm max}$ and ${\cal L}_{0}$ denote the maximum likelihood 
value and likelihood value obtained assuming zero signal events, respectively.
Here the likelihood function from the fit is convolved with a Gaussian
systematic error function in order to include the systematic uncertainty
in the signal yield.


\section{Results from the \boldmath$\Upsilon(5S)$ Engineering Run}

We use a data sample of $1.86~\textrm{fb}^{-1}$ taken at the $\Upsilon(5S)$ energy of
$\sim 10869$ MeV. The experimental conditions of data taking at $\Upsilon(5S)$ are identical
to that for $\Upsilon(4S)$ or continuum running. The data sample of $3.67~\textrm{fb}^{-1}$ taken in
the continuum at an energy of $60$ MeV below the $\Upsilon(4S)$ was also used in this analysis for
comparison.

The $B_{s}$ mesons are produced at the $\Upsilon(5S)$ through
the intermediate $B_{s}\bar{B}_{s}$, $B_{s}^{*}\bar{B}_{s}$, $B_{s}\bar{B}_{s}^{*}$ or
$B_{s}^{*}\bar{B}_{s}^{*}$ pair production channels, where $B_{s}^{*}$ decays to $B_{s}\gamma$.
These intermediate channels can be distinguished kinematically and
their production ratios can be obtained from the reconstruction of exclusive $B_{s}$ decays.
To improve the statistical significance of our exclusive $B_{s}$ signal, we combined the
six modes 
$B_{s}\rightarrow D_{s}^{+}\pi^-$, 
$B_{s}\rightarrow D_{s}^{*+}\pi^-$,
$B_{s}\rightarrow D_{s}^{+}\rho^-$, 
$B_{s}\rightarrow D_{s}^{*+}\rho^-$,
$B_{s}\rightarrow J/\psi\phi$ and 
$B_{s}\rightarrow J/\psi\eta$
,which have large reconstruction efficiencies and are described by
unsuppressed conventional tree diagrams.

Six conventional $B_{s}$ decays to $D_{s}^{+}\pi^{-}$, $D_{s}^{+}\rho^{-}$, 
$D_{s}^{*+}\pi^{-}$, $D_{s}^{*+}\rho^{-}$, $J/\psi\phi$ and $J/\psi\eta$
final states and four rare $B_{s}$ decays to $K^+K^-$, $\phi\gamma$, $\gamma\gamma$ and
$D_{s}^{(*)+}D_{s}^{(*)-}$ final states are reconstructed.
The signals can be observed using two variables:
the energy difference $\Delta E = E_{B_{s}}^{\rm CM} - E_{\rm beam}^{\rm CM}$
and beam-constrained mass $M_{\rm bc} = \sqrt{(E_{\rm beam}^{\rm CM})^{2}
- (p_{B_{s}}^{\rm CM})^2 }$; 
$E_{B_{s}}^{\rm CM}$ and $p_{B_{s}}^{\rm CM}$
are the energy and momentum of the $B_{s}$ candidate in the CM 
system and $E_{\rm beam}^{\rm CM}$ is the CM beam energy. 
The $B_{s}$ mesons can be produced at the $\Upsilon(5S)$ energy via the intermediate
$e^+e^-\rightarrow B_{s}^{(*)}\bar{B}_{s}^{(*)}$ channels, with 
$B_{s}^{*}\rightarrow B_{s}\gamma$.
The $B_{s}$ signal regions in $M_{\rm bc}$ and $\Delta E$ are separated for 
different intermediate channels. 

\begin{figure}
\begin{center}
 \includegraphics[width=80mm]{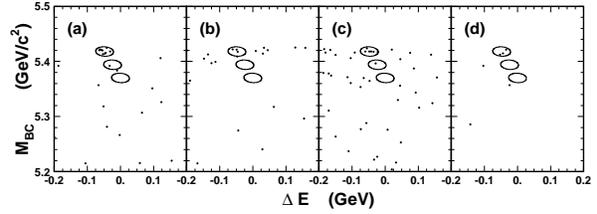}
\caption{The $M_{\rm bc}$ and $\Delta E$ scatter plot for $B_{s}\rightarrow D_{s}^+\pi^-$(a),
$B_{s}\rightarrow D_{s}^{*+}\pi^-$(b) and $B_{s}\rightarrow D_{s}^{(*)+}\rho^-$(c) decay modes are shown,
where $D_{s}^{+}$ meson is reconstructed in the $D_{s}^{+}\rightarrow\phi\pi^{+}$, $D_{s}^{+}\rightarrow \bar{K}^{*0} K^{+}$
and $D_{s}^{+}\rightarrow K_{S}^{0} K^{+}$ decay modes. Also $M_{\rm bc}$ and $\Delta E$ scatter plot (d) is shown for
the $B_{s}\rightarrow J/\psi\phi$ and $B_{s}\rightarrow J/\psi\eta$.}
    \label{fig:mbc_deltae}
\end{center}   
\end{figure}

After all selections, the dominant background is from $e^+e^-\rightarrow q\bar{q}$
continuum events ($q = u,~d,~s,~\mbox{or}~c$). 
The distribution of data in $M_{\rm bc}$ and $\Delta E$ for the 
$B_{s}\rightarrow D_{s}^+\pi^-$ decay mode is shown in Figure~\ref{fig:mbc_deltae}a. 
Three $D_{s}^+$ decay modes, $\phi\pi^+$, $\bar{K}^{*0}K^+$ and $K_{S}^{0}K^+$, are used
to reconstruct $B_{s}$ candidates.
Nine events are observed within the $B_{s}$ signal ellipsoidal region corresponding to 
$B_{s}^{*}\bar{B}_{s}^{*}$ pair production channel. 
Only one event is observed in the $B_{s}$ signal region for 
$B_{s}^{*}\bar{B}_{s} + B_{s}\bar{B}_{s}^{*}$ channels, 
and no events are observed for $B_{s}\bar{B}_{s}$ channel. 
Background outside the signal regions is small and corresponds to $0.1$ event 
for any of three signal regions.
The inclusive studies at the $\Upsilon(5S)$ found that $92,000\pm 7,900\pm 23,500$
$B_{s}^{(*)}\bar{B}_{s}^{(*)}$ pairs are contained within that $1.86~\mbox{fb}^{-1}$~
$\Upsilon(5S)$ data sample. 
Using this value, we measure the branching fraction to be 
${\cal B}(B_{s}\rightarrow D_{s}^+\pi^-) = (0.65\pm  0.21\pm  0.19)\%$.

The $M_{\rm bc}$ and $\Delta E$ scatter plots are also obtained for the 
$B_{s}\rightarrow D_{s}^{*+}\pi^-$ (Figure~\ref{fig:mbc_deltae}b)
and $B_{s}\rightarrow D_{s}^{(*)+}\rho^-$ (Figure~\ref{fig:mbc_deltae}c) decay modes. 
Again, three $D_{s}^+$ decay modes, $\phi\pi^+$, $\bar{K}^{*0}K^+$ and $K_{S}^{0}K^+$, are used
to reconstruct $B_{s}$ candidates.
The numbers of events within the $B_{s}$ signal region for the $B_{s}^{*}\bar{B}_{s}^{*}$
pair production channel are $4$ for $B_{s}\rightarrow D_{s}^{*+}\pi^-$ decay and 
$7$ for $B_{s}\rightarrow D_{s}^{(*)+}\rho^-$ decay. 

The scatter plot in $M_{\rm bc}$ and $\Delta E$ for the $B_{s}\rightarrow J/\psi\phi$
and $B_{s}\rightarrow J/\psi\eta$ decays is shown in Figure~\ref{fig:mbc_deltae}d. 
One of the observed $B_{s}\rightarrow J/\psi\phi$ candidates is reconstructed 
in the $J/\psi\rightarrow\mu^{+}\mu^{-}$ mode and one in the 
$J/\psi\rightarrow e^{+}e^{-}$ mode. 
These two candidates correspond roughly to a branching fraction of
$\sim 1\times 10^{-3}$, in agreement with expectations.

\begin{figure}
\begin{center}
 \includegraphics[width=55mm]{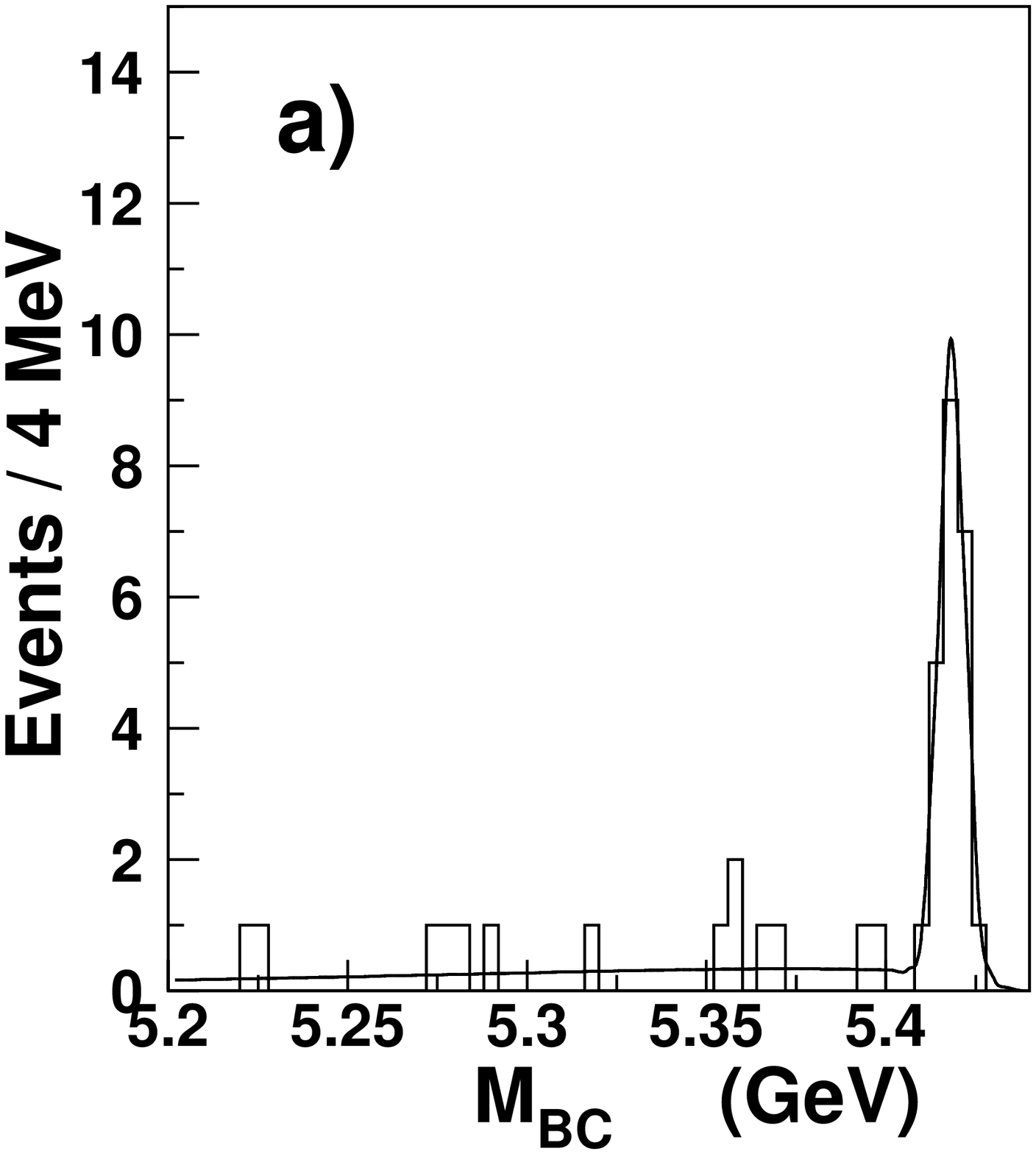}
 \includegraphics[width=55mm]{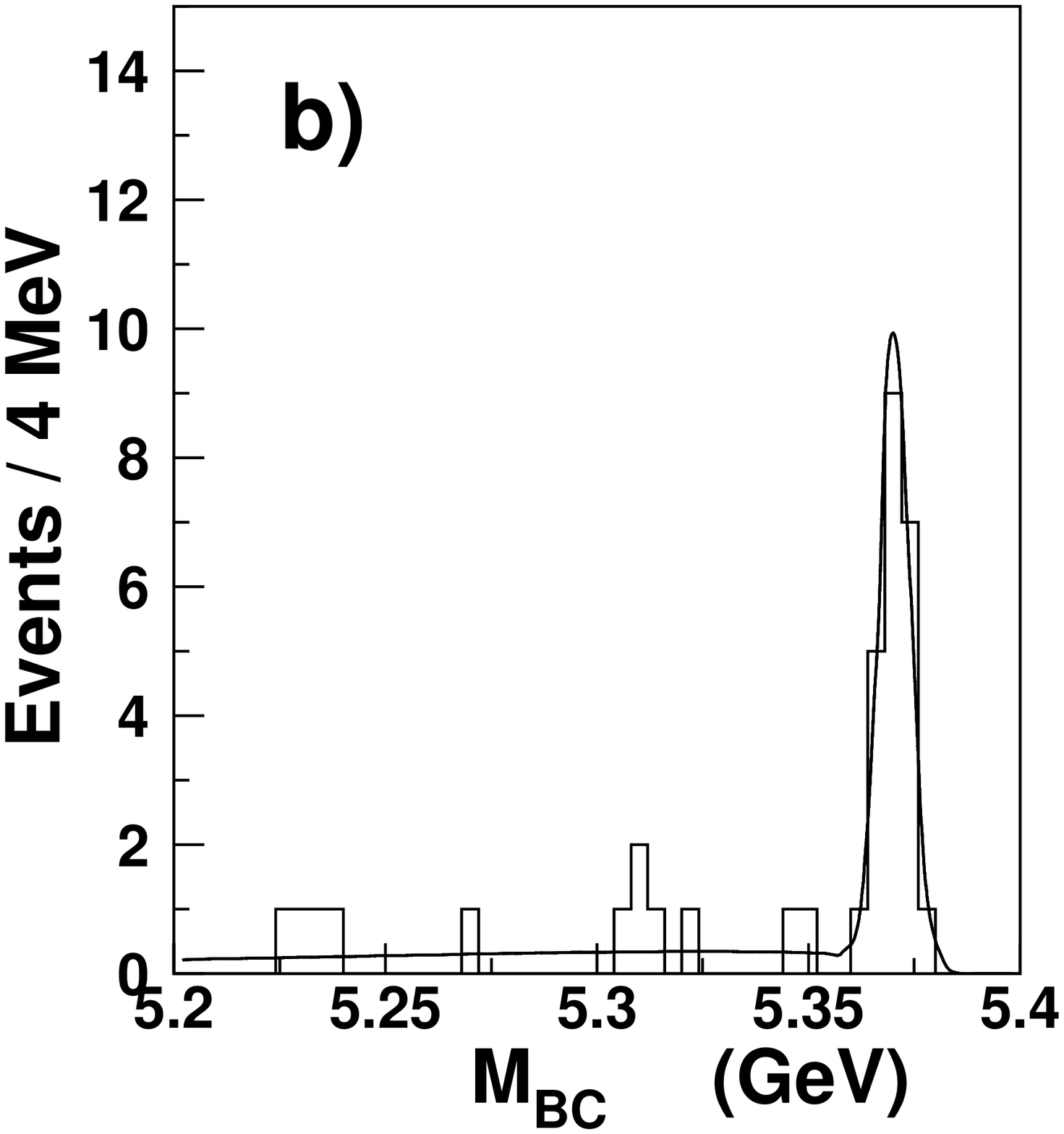}
\caption{The $B_{s}^{*}$ (a) and $B_{s}$ (b) mass distributions for events within the
  $-0.08 < \Delta E < -0.02$ MeV interval, corresponding to the $B_{s}^{*}\bar{B}_{s}^{*}$ channel.}
    \label{fig:mbc}
\end{center}   
\end{figure}

The $B_{s}$ and $B_{s}^{*}$ masses can be extracted from the $M_{\rm bc}$ fits in the 
$B_{s}^{*}\bar{B}_{s}^{*}$ channel.
The $M_{\rm bc}$ distribution for this channel (Figure~\ref{fig:mbc}a) is obtained choosing candidates
within the $-0.08 < \Delta E < -0.02$ MeV range. 
The distribution, shown in Figure~\ref{fig:mbc}a, is fitted by the
sum of a Gaussian to describe the signal and the ARGUS function to describe the background. 
The fit yields the mass value 
$M(B_{s}^{*}) = 5418\pm  1~\mbox{MeV}/c^{2}$.
The observed width of the $B_{s}$ signal is $3.6\pm 0.6~\mbox{MeV}/c^{2}$ and 
agrees with the value obtained from the MC simulation.
Using events from the $B_{s}^{*}\bar{B}_{s}^{*}$ channel we can obtain also the 
$B_{s}$ mass (Figure~\ref{fig:mbc}b).
The distribution shown in Figure~\ref{fig:mbc}b is fitted to the sum of a Gaussian 
and the ARGUS function.
The fit yields the $B_{s}$ mass 
$M(B_{s}) = 5370\pm  1\pm  3~\mbox{MeV}/c^{2}$ and width 
$\sigma(B_{s}) = 3.6\pm  0.6~\mbox{MeV}/c^{2}$.

Additionally, we searched for several $B_{s}$ rare decays for the first time: 
the penguin decay $B_{s}\rightarrow K^+K^-$, the electromagnetic penguin decay 
$B_{s}\rightarrow\phi\gamma$, and the intrinsic penguin decay $B_{s}\rightarrow\gamma\gamma$. 
We also searched for the tree decay
$B_{s}\rightarrow D_{s}^{(*)+}D_{s}^{(*)-}$, which is not yet observed and 
is of special interest because the $D_{s}^{(*)+}D_{s}^{(*)-}$ states are expected to be 
dominantly CP eigenstates. Although the
branching fractions for these decays are expected to be too low for observation in
this analysis, we obtained upper limits (Table~\ref{tab:upper_limit}).

\begin{table}
  \renewcommand{\baselinestretch}{1.3}
    \begin{tabular}{|c|c|c|c|c|} \hline
Decay mode  & Yield  & Background & Eff.  & upper limit\\
            &events  &events       & (\%)         &($\times 10^{-4}$)\\
\hline
$B_{s}\rightarrow K^+K^-$                 &$2$    &$0.14$   &$9.5$     &$3.4$\\\hline
$B_{s}\rightarrow\phi\gamma$              &$1$    &$0.15$   &$5.9$     &$4.1$\\\hline
$B_{s}\rightarrow\gamma\gamma$            &$0$    &$0.5$    &$20.0$    &$0.56$\\\hline
$B_{s}\rightarrow D_{s}^{+}D_{s}^{-}$     &$0$    &$0.02$   &$0.020$   &$710$\\\hline
$B_{s}\rightarrow D_{s}^{*+}D_{s}^{-}$    &$1$    &$0.01$   &$0.0090$   &$1270$\\\hline
$B_{s}\rightarrow D_{s}^{*+}D_{s}^{*-}$   &$0$    &$<0.01$   &$0.0052$   &$2730$\\\hline
    \end{tabular}
    \caption{The number of events in the signal region, estimated background events and upper limits on the $90\%$ confidence
    level for $B_{s}\rightarrow K^+K^-$, $B_{s}\rightarrow\phi\gamma$, $B_{s}\rightarrow\gamma\gamma$ and 
    $B_{s}\rightarrow D_{s}^{(*)+}D_{s}^{(*)-}$ decay modes.}
   \label{tab:upper_limit}
\end{table}

\bigskip 
\begin{acknowledgments}
The author wish to thank the KEKB accelerator group for the excellent operation of the KEKB
accelerator.
\end{acknowledgments}

\bigskip 


\end{document}